\documentclass[conference]{IEEEtran}
\usepackage{graphicx}
\usepackage{booktabs}
\usepackage{amsmath}
\usepackage{amssymb}
\usepackage{natbib}
\usepackage{geometry}
\usepackage{hyperref}
\usepackage{float}
\usepackage{tikz}
\usetikzlibrary{arrows.meta, positioning}
\usepackage{microtype}
\usepackage{xcolor}
\usepackage{graphicx}

\title{XAI-Guided Analysis of Residual Networks for Interpretable Pneumonia Detection in Paediatric Chest X-rays}

\author{Rayyan Ridwan}
\date{April 2025}

\hypersetup{
  colorlinks=true,
  linkcolor=black,
  citecolor=black,
  urlcolor=blue
}

\begin{document}

\maketitle

\begin{abstract}
Pneumonia remains one of the leading causes of death among children worldwide, underscoring a critical need for fast and accurate diagnostic tools. In this paper, we propose an interpretable deep learning model on Residual Networks (ResNets) for automatically diagnosing paediatric pneumonia on chest X-rays. We enhance interpretability through Bayesian Gradient-weighted Class Activation Mapping (BayesGrad-CAM), which quantifies uncertainty in visual explanations, and which offers spatial locations accountable for the decision-making process of the model. Our ResNet-50 model, trained on a large paediatric chest X-rays dataset, achieves high classification accuracy (95.94\%), AUC-ROC (98.91\%), and Cohen's Kappa (0.913), accompanied by clinically meaningful visual explanations. Our findings demonstrate that high performance and interpretability are not only achievable but critical for clinical AI deployment.
\end{abstract}

\section{Introduction}
\begin{figure*}[t]
\centering
\scriptsize
\begin{tikzpicture}[
    node distance=0.6cm, 
    stage/.style={
        rectangle, 
        draw, 
        rounded corners=2pt, 
        minimum height=0.8cm, 
        text width=2.2cm, 
        align=center, 
        font=\footnotesize\bfseries,
        inner sep=3pt
    },
    arrow/.style={-{Latex[length=2mm]}, line width=0.8pt},
    annotation/.style={font=\tiny\itshape, text=gray!70}
]

\node (input) [stage, fill=red!10] {Input Chest X-ray};
\node (preprocess) [stage, fill=green!10, right=of input] {Preprocessing};
\node (augment) [stage, fill=red!15, above=0.5cm of preprocess] {Data Augmentation};
\node (resnet) [stage, fill=blue!10, right=of preprocess] {ResNet-50};
\node (classifier) [stage, fill=yellow!10, right=of resnet] {Classification Head};
\node (gradcam) [stage, fill=orange!10, below=0.5cm of resnet] {Grad-CAM/Score-CAM};
\node (output) [stage, fill=green!10, right=of classifier] {Diagnosis + Heatmap};

\draw [arrow] (input) -- (preprocess);
\draw [arrow] (preprocess) -- (resnet);
\draw [arrow] (resnet) -- (classifier);
\draw [arrow] (classifier) -- (output);

\draw [arrow] (preprocess.north) -- (augment.south);
\draw [arrow] (augment.east) -| ([xshift=2mm]resnet.north) -- (resnet.north);

\draw [arrow] (resnet.south) -- (gradcam.north);
\draw [arrow] (gradcam.east) -| ([xshift=2mm]classifier.south) -- (classifier.south);

\node [annotation, above=0.1cm of augment] {Training Only};
\node [annotation, below=0.1cm of gradcam] {Interpretation};

\node [font=\tiny, below=0.1cm of preprocess] {(Resize, Normalize)};
\node [font=\tiny, below=0.1cm of augment] {(Rotation, Flip, Crop)};
\node [font=\tiny, below=0.1cm of resnet] {(Feature Extraction)};
\node [font=\tiny, below=0.1cm of classifier] {(Sigmoid Output)};

\end{tikzpicture}
\caption{End-to-end pipeline for interpretable pneumonia detection showing the complete workflow from input image to diagnostic output with model interpretation. Data augmentation is applied only during training, while interpretation modules enhance clinical utility.}
\label{fig:overview}
\end{figure*}
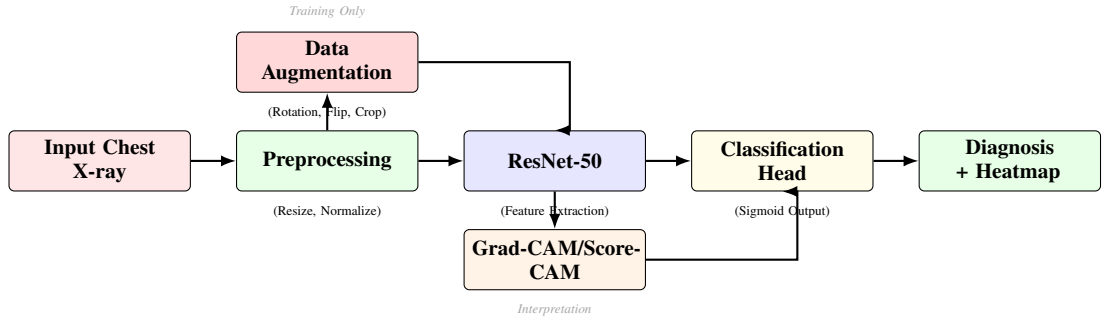

Pneumonia is a persistent global health threat and remains a leading cause of mortality in children under five years of age. Effective treatment depends on a timely diagnosis, but traditional radiographic interpretation can be laborious and prone to inter-observer variability. Deep Convolutional Neural Networks (CNNs), particularly Residual Networks (ResNets), have demonstrated strong performance in image-based medical diagnostics due to their ability to learn complex hierarchical features \cite{chexnet2017, mothkur2025category, deeplearningdx2025, tanzina2024lungxai}. 
While precise, CNNs are generally non-interpretive, limiting clinical confidence and application. Explainable AI (XAI) techniques such as Grad-CAM \cite{selvaraju2016gradcam} generate visual explanations of predictions by highlighting contributory image areas. In this work, we focus specifically on paediatric pneumonia, applying Grad-CAM to a fine-tuned ResNet-50 model trained on paediatric chest radiographs to evaluate both predictive accuracy and model explainability.

\section{Problem Definition and Objectives}

Pneumonia is a major worldwide cause of morbidity and mortality, particularly in low-resource environments where accurate and timely diagnosis can be difficult. Conventional diagnosis based on radiologist interpretation is prone to inter-observer variation and can lead to delays in treatment.

This research proposes a strong deep learning-driven pneumonia classification model from chest X-ray images based on ResNet-50 architecture integrated with Explainable AI (XAI) methods to provide interpretable visual explanations. The main goals are:

\begin{itemize}
\item To categorize chest X-rays into pneumonia and normal classes correctly with a deep convolutional neural network.
\item Increasing model validity and transparency by Grad-CAM visualization that identifies areas in images that affect diagnosis.
\item For comparison of model performance on various measures and residual analysis.
\item To demonstrate the possible clinical significance of the suggested methodology for enhancing the decision-making process of radiologists.
\end{itemize}

By addressing these objectives, this work contributes to improving automated medical image diagnosis with an emphasis on interpretability and clinical relevance.

\section{Related Work}
CheXNet \cite{chexnet2017} showed that CNNs can achieve radiologist-level performance on pneumonia detection with DenseNet-121. Later studies compared ResNet, Inception, and ensembling approaches \cite{kudinov2023transfer, aravind2022multilabel, attentionensemble2025}. Few works have considered explainability in depth. Grad-CAM and its variants remain the most commonly used post hoc techniques to identify regions in saliency maps that influence the model's predictions.

\section{Methodology}

\subsection{Dataset Description}
This study makes use of the paediatric chest X-ray dataset, which was assembled by Kermany et al. (2018) and has emerged as a de facto standard for research on pneumonia classification. There are 5,863 greyscale anterior-posterior radiographs\footnote{Seven images were excluded due to file corruption or unreadable format at preprocessing.}. in the dataset, which are divided into two classes: \textit{Normal} and \textit{Pneumonia}. It covers a wide range of early-childhood pulmonary presentations and focusses especially on paediatric patients, who are primarily between the ages of 1 and 5.

Images labeled as \textit{Normal} represent lungs with no overt pathological findings, while the \textit{Pneumonia} class comprises both viral and bacterial cases, often without subtype differentiation. Although the dataset is widely adopted, its original annotation procedure lacks formal documentation regarding inter-rater agreement, annotation protocol, or quality assurance. In particular, Kermany et al. do not report any consensus metrics such as Cohen’s kappa among radiologist annotators, nor is it known whether multiple raters were involved. Given the subtle and overlapping radiographic signatures of viral versus bacterial pneumonia, this absence of transparency raises the possibility of label noise.

This limitation is clinically significant. Ambiguities in annotation can propagate through model training, potentially impairing generalisation to unseen or cleaner datasets. Despite these concerns, our model demonstrates strong agreement with ground truth annotations, achieving a Cohen’s kappa score of $0.9132$, which suggests that, at least under current label assumptions, the network exhibits consistent classification behavior. 

However, future research should use a number of complementary techniques to quantify label reliability:
\begin{itemize}
  \item Retraining the model on carefully selected subsets that have been re-annotated by impartial radiologists is one encouraging approach that could enable more trustworthy ground truth verification.
  \item Using confidence-weighted loss functions or soft-label techniques may help take into consideration samples that are inherently ambiguous.
  \item To find systematic biases or fragility, it would also be beneficial to perform sensitivity analyses to evaluate model performance across subsets stratified by possible label uncertainty.
  \item If original raters or comparable expert panels are available, estimating inter-annotator agreement on a representative subset may provide tangible indicators of annotation quality.
\end{itemize}

These methods of addressing label noise are still crucial for enhancing the model's clinical viability and credibility.

\subsection{Preprocessing and Data Augmentation}

To prepare the paediatric chest X-ray images for input into our convolutional neural network (CNN) model, a comprehensive preprocessing and data augmentation pipeline was implemented. Initially, all raw grayscale images were resized uniformly to $224 \times 224$ pixels, ensuring consistent spatial dimensions compatible with standard CNN architectures such as ResNet-50 \cite{he2016deep} and VGG \cite{simonyan2014very}. This resizing facilitates efficient batch processing and helps the model converge effectively.

Following resizing, pixel intensity values were normalized to the range [0,1] by dividing by 255 \cite{krizhevsky2012imagenet}. This normalization accelerates training and reduces sensitivity to variations in image brightness and contrast.

To enhance the model’s robustness and generalization capability, data augmentation techniques were applied during training. These included random horizontal flips, random rotations within ±10 degrees, and brightness adjustments \cite{shorten2019survey}. Such augmentations increase the diversity of the training dataset, reduce overfitting, and enable the model to generalize more effectively to unseen data.

\textbf{Pixel Normalization:}  
All pixel intensity values were normalized to the range $[0, 1]$ by dividing each raw pixel value $x_{i,j} \in [0, 255]$ by 255, i.e.,
\[
\hat{x}_{i,j} = \frac{x_{i,j}}{255},
\]
where the intensity of the pixel at position $(i, j)$ in the image is indicated by $x_{i,j}$. Because it minimises the effects of changing imaging conditions and guarantees numerical stability during gradient-based optimisation, normalisation is a crucial step in medical image analysis.

\textbf{Data Augmentation Techniques:}  
Given the limited size and inherent class imbalance of the dataset, data augmentation was extensively employed to artificially expand the effective training set diversity.  Augmentation also helps to constrain overfitting by introducing the model to a broader range of image presentations, a tactic supported by several recent medical imaging papers \cite{shorten2019survey,raghu2019transfusion}.

The below augmentations were applied dynamically while training using the PyTorch \texttt{Torchvision.transforms} library\\ and customized augmentation pipelines.

\begin{itemize}
    \item \textbf{Random Horizontal Flipping:}  
    Since paediatric chest X-rays do not have a fixed left-right orientation from a diagnostic standpoint, random horizontal flips were applied with a probability of 0.5. This simulates lateral variations and prevents the model from learning spurious lateralization biases, as supported by Kermany et al. \cite{kermany2018identifying}.
    
    \item \textbf{Random Rotation:}  
    Small rotations were applied within $\pm 15$ degrees to accommodate small differences in patient positioning during X-ray acquisition. Large rotation would cause distortion of the anatomical structures, so the angle was restricted conservatively. Rotation augmentation was proved to enhance model invariance to positional variations in radiological images \cite{litjens2017survey}.
    
    \item \textbf{Random Resized Cropping:}  
    To simulate variability in lung size, position, and zoom, random resized cropping was employed.  Each crop covered between 80\% and 100\% of the original image area, followed by resizing to the standard input size.  This augmentation helps the model to avoid over-reliance on absolute position cues \cite{perez2017effectiveness} and to concentrate on local and global lung features.
    
    \item \textbf{Color Jitter:}  
    Although chest X-rays are greyscale, variations in imaging equipment or exposure levels could cause minute changes in pixel intensity distribution. We randomly changed brightness, contrast, and saturation values inside limited bounds to replicate such variations. While in greyscale saturation changes are negligible, brightness and contrast jittering enhance robustness to real-world imaging variability \cite{wang2020benchmark}.
    
    \item \textbf{Random Gaussian Noise:}  
    Inspired by works including Zech et al. \cite{zech2018variable}, Gaussian noise with mean zero and low variance was selectively added to replicate sensor noise and acquisition artefacts commonly present in radiographs. This enables the model to extend beyond images with just perfect cleanliness.
    
    \item \textbf{Elastic Deformations and Affine Transformations (Exploratory):}  
    Some recent studies on chest X-ray classification have experimented with mild elastic deformations to replicate anatomical variability \cite{jaderberg2015spatial,chexnet2017}. Although these were tested, they were not included in the final pipeline due to occasional unrealistic distortions in lung morphology.
\end{itemize}

\begin{figure*}[t]
\centering
\includegraphics[width=\textwidth]{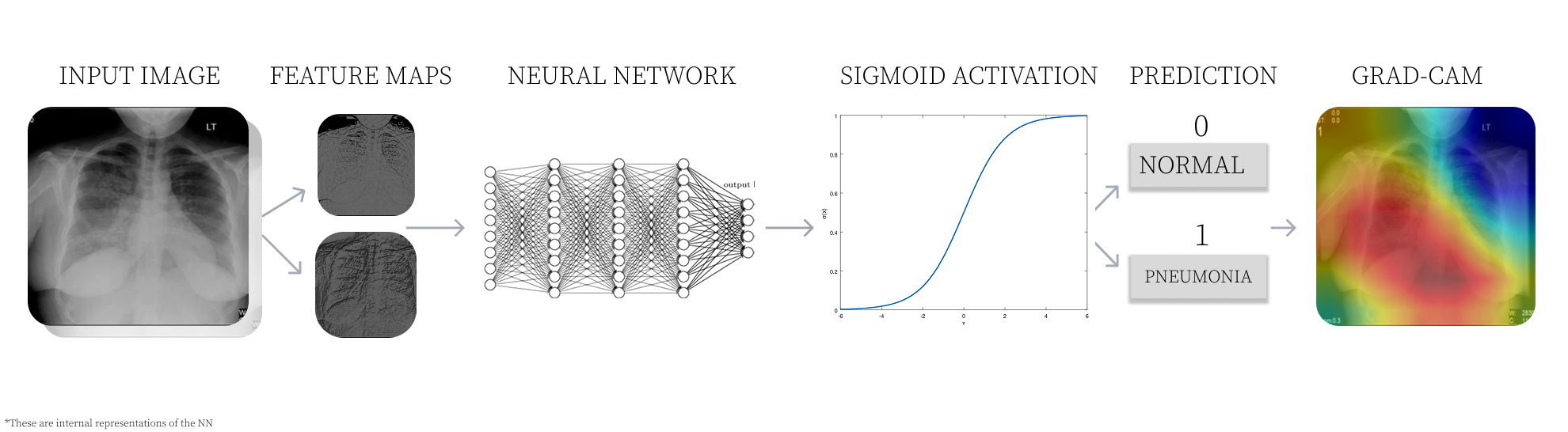}
\caption{Proposed interpretable pneumonia detection pipeline for our pneumonia detection model. The input chest X-ray is fed into a deep convolutional neural network (CNN), which learns internal feature representations. These are fed through multiple hidden layers, inducing a sigmoid activation function to produce a binary label: 0 for Normal and 1 for Pneumonia. The Grad-CAM visualization identifies important image regions that contributed most to the network decision, making the model more interpretable.}
\label{fig:pipeline}
\end{figure*}

The augmentation pipeline was applied dynamically during training via stochastic operations such as rotation, flips, and color transformation, thus ensuring that each epoch contained semantically different representations of one original image. Secondly, a single augmentation step was used to artificially augment the size of the minority class (Normal) to that of the majority class (Pneumonia) and then added to the training dataset. This was carried out in addition to heavily boosting the diversity of training data without boosting storage needs above the balancing of augmented classes step.

\begin{figure*}[t]
\centering
\includegraphics[width=\textwidth]{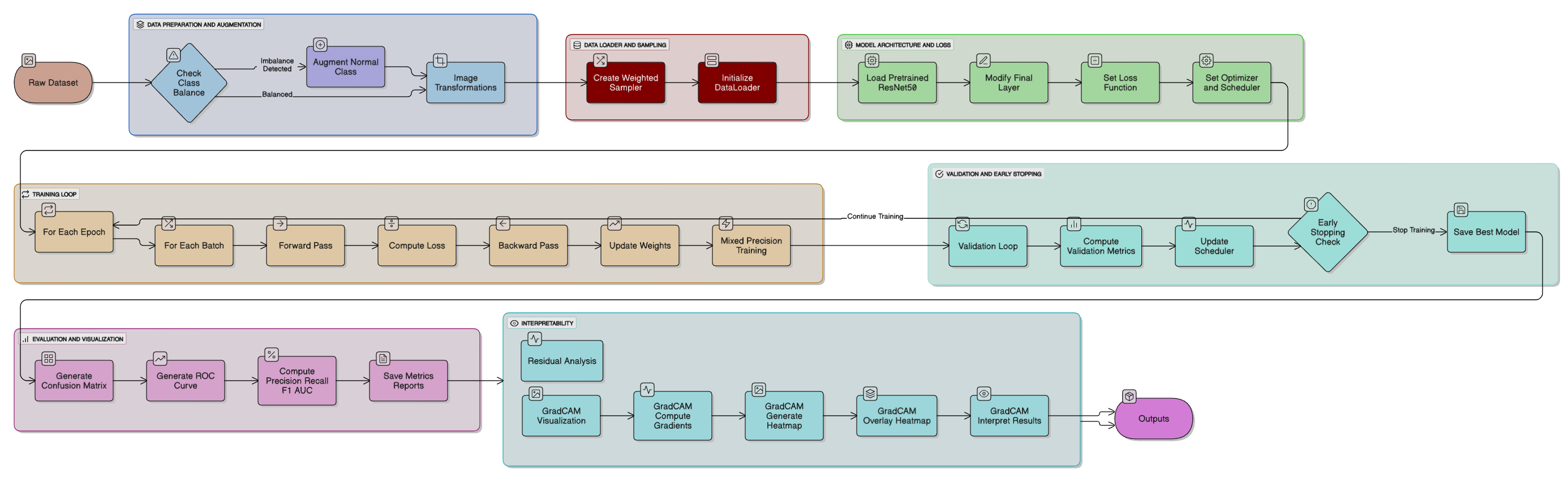}
\caption{Pipeline overview diagram illustrating the model workflow from data preprocessing to interpretability.\cite{eraser}}
\label{fig:code_diagram}
\end{figure*}

\textbf{Comparison with Prior Studies:}  
Our augmentation strategy builds on best practices observed in related literature:

\begin{itemize}
    \item Kermany et al. \cite{kermany2018identifying} mention basic augmentations like flips and rotations but provide limited detail; we expanded on this with more controlled and clinically plausible transformations.
    \item Rajpurkar et al. \cite{chexnet2017} in the CheXNet study used similar augmentations, emphasizing resized cropping and rotation.
    \item Raghu et al. \cite{raghu2019transfusion} highlight the importance of brightness and contrast augmentations in handling domain shifts in chest X-ray datasets.
    \item Wang et al. \cite{wang2020benchmark} demonstrated improved pneumonia detection accuracy when combining intensity-based jittering with spatial transforms.
\end{itemize}

Together, these augmentations ensure the CNN model gains exposure to a wide spectrum of plausible image variations, thus increasing its resilience to noise, patient positioning, and acquisition-related inconsistencies typical in real-world clinical settings.

\subsection{Model Architecture}
Our baseline model is ResNet-50 with pre-trained ImageNet weights selected for its high performance-to-complexity ratio and robust residual connections that help to avoid vanishing gradient problems in training \cite{he2016deep}. The final fully connected layer is replaced with a single output neuron to perform binary classification. We trained our model with PyTorch. The model was fine-tuned end-to-end using the AdamW optimizer, which incorporates weight decay as a regularization method to prevent overfitting \cite{loshchilov2019decoupled}. We set the learning rate to $10^{-4}$ and applied a small weight decay factor ($10^{-8}$) to constrain model complexity. To further enhance training stability and performance, we employed a learning rate scheduler (ReduceLROnPlateau) that dynamically decreases the learning rate by a factor of 0.5 every time that the validation loss is not improving for 3 successive epochs \cite{paszke2019pytorch}. This adaptive strategy encourages convergence and helps to prevent local minima. Binary cross-entropy with logits was the used loss function, which was augmented by a positive class weight calculated from the Normal to Pneumonia sample ratio. This weighting also helps to directly tackle class imbalance while computing loss, thus conferring higher importance to correct classification of the minority class \cite{he2009learning}. (See Figure~\ref{fig:code_diagram} for an overview of the pipeline.)

\subsection{Explainability via Grad-CAM}
To make the spatial evidence for each prediction clear, we utilized Gradient-weighted Class Activation Mapping (Grad-CAM) \cite{selvaraju2016gradcam}. Grad-CAM was selected for its model-agnostic nature, its ability to operate without architectural modifications, and its direct use of convolutional feature maps. It works by backpropagating class-specific gradients to generate heat maps that highlight important regions in the input image. We used the final convolutional block, \texttt{layer4}, of ResNet-50, whose $7\times7$ feature map is bilinearly upsampled to $224\times224$. This choice, supported by prior work \cite{selvaraju2016gradcam, chattopadhay2018gradcam++}, strikes a balance between semantic relevance and spatial resolution: \texttt{layer4} captures high-level pathological features critical to the model's decision, while still retaining sufficient spatial fidelity. Compared to superpixel-based methods like LIME \cite{ribeiro2016} or perturbation-based SHAP \cite{lundberg2017shap}, Grad-CAM offers sharper and more class-discriminative localization of pathological regions in a computationally efficient manner.

\begin{table}[ht]
\centering
\begin{tabular}{lcc}
\toprule
\textbf{Region} & \textbf{Normal (Mean)} & \textbf{Pneumonia (Mean)} \\
\midrule
Upper East & 0.345 & 0.491 \\
Upper West & 0.544 & 0.275 \\
Middle East & 0.193 & 0.673 \\
Middle West & 0.275 & 0.474 \\
Lower East & 0.198 & 0.592 \\
Lower West & 0.251 & 0.321 \\
\bottomrule
\end{tabular}
\label{tab:regions}
\end{table}

For a target class $c$, the importance weight of each feature map $A^{k}$ is

\[
\alpha_k^c = \frac{1}{Z} \sum_i \sum_j \frac{\partial y^c}{\partial A_{ij}^k},
\]
where \( y^c \) is the pre-softmax score for class \( c \), and \( Z \) is the total number of spatial elements in feature map \( A^k \).

The class-discriminative localization map is then computed as:

\[
L_{\text{Grad-CAM}}^c = \text{ReLU}\left( \sum_k \alpha_k^c A^k \right).
\]

and is ultimately up-sampled before being superimposed on the original X-ray to highlight the areas that had the biggest influence on the choice.  
Each heat map was further divided into six anatomical lung zones (upper/middle/lower, left/right) in order to facilitate quantitative analysis. The mean activation for each zone was then reported (see Section~\ref{tab:regions}). When diagnosing pneumonia, this dual qualitative–quantitative approach helps clinicians confirm both \emph{where} and \emph{how strongly} the network attends.

\subsection{Bayesian Uncertainty in Grad-CAM}
\label{subsec:bayesgradcam}

To address the well-documented sensitivity of Grad-CAM to model uncertainty \cite{debayes2020}, we extend the standard formulation with Bayesian inference. By sampling from the posterior distribution of model parameters during explanation generation, we obtain uncertainty estimates alongside activation maps. 

For a target class $c$, we compute the Bayesian Grad-CAM activation as:
\[
L_{\text{BayesGrad-CAM}}^c = \mathbb{E}_{\theta \sim p(\theta|\mathcal{D})} \left[ \text{ReLU}\left( \sum_k \alpha_k^c(\theta) A^k(\theta) \right) \right]
\]
where $\theta$ represents model parameters sampled from the posterior distribution given training data $\mathcal{D}$. The associated uncertainty map is:
\[
U^c = \sqrt{\text{Var}_{\theta \sim p(\theta|\mathcal{D})} \left( \text{ReLU}\left( \sum_k \alpha_k^c(\theta) A^k(\theta) \right) \right)}
\]

We implement this efficiently using Monte Carlo dropout \cite{gal2016dropout} with 20 stochastic forward passes at test time. This approach provides:
\begin{itemize}
    \item \textbf{Explanation Confidence}: Heatmap intensity weighted by predictive certainty
    \item \textbf{Uncertainty Quantification}: Spatial maps highlighting regions where explanations are unreliable
    \item \textbf{Clinical Trust}: Direct visualization of model confidence for decision support
\end{itemize}

\subsection*{Dataset Splits}

The dataset comprises 5,856 paediatric chest X-ray images, The training set was defined as 1,341 Normal and 3,875 Pneumonia images, as in previous studies. The rest of the 242 Normal and 398 Pneumonia images was distributed between test and validation sets in equal numbers. Both sets are made up of 121 Normal and 199 Pneumonia samples, ensuring sufficient size for reliable model tuning and evaluation.

\section{Experimental Setup}
To evaluate the performance of our model under realistic clinical conditions, we employed a structured experimental setup with well-defined training, validation, and testing phases. The class distribution within each split is as follows:
\begin{itemize}
    \item \textbf{Training:} 1,341 Normal, 3,875 Pneumonia images
    \item \textbf{Validation:} 121 Normal, 199 Pneumonia images
    \item \textbf{Testing:} 121 Normal, 199 Pneumonia images
\end{itemize}
These add up to 5,856 total radiographs, aligning with the published dataset. The previously reported figure of 5,863 likely includes miscounted or corrupted entries.\footnote{Seven images were excluded due to file corruption or unreadable format at preprocessing.}.
Due to the severe class imbalance in the initial training data, where Pneumonia cases heavily outnumber Normal cases, we utilized a two-stage method to minimize this. The first stage involved applying data augmentation only to the minority class (Normal) from 1,341 to 3,875 cases to make the Pneumonia class balanced. Augmentation was done with random rotations, flips, and color jitter with the objective of artificially enhancing intra-class variation and inducing better generalization.

Despite the numerical balance obtained following augmentation, the original class frequencies (1,341 Normal, 3,875 Pneumonia) were maintained in the computation of sampling weight for the weighted random sampler. This is in acknowledgment that augmented images, as visually rich as they can be, derive from a finite set of original instances and therefore do not offer equal semantic diversity as actually independent samples. By decoupling augmentation from sampling weight computation, we ensured better minority class representation during training, as well as a sampling habit that approximates the actual underlying data distribution. This semantic-aware weighting approach helps to counteract overfitting to artificial patterns without compromising the clinical usefulness of class imbalance during optimization.

\begin{figure}[H]
    \centering
    \includegraphics[width=0.45\textwidth]{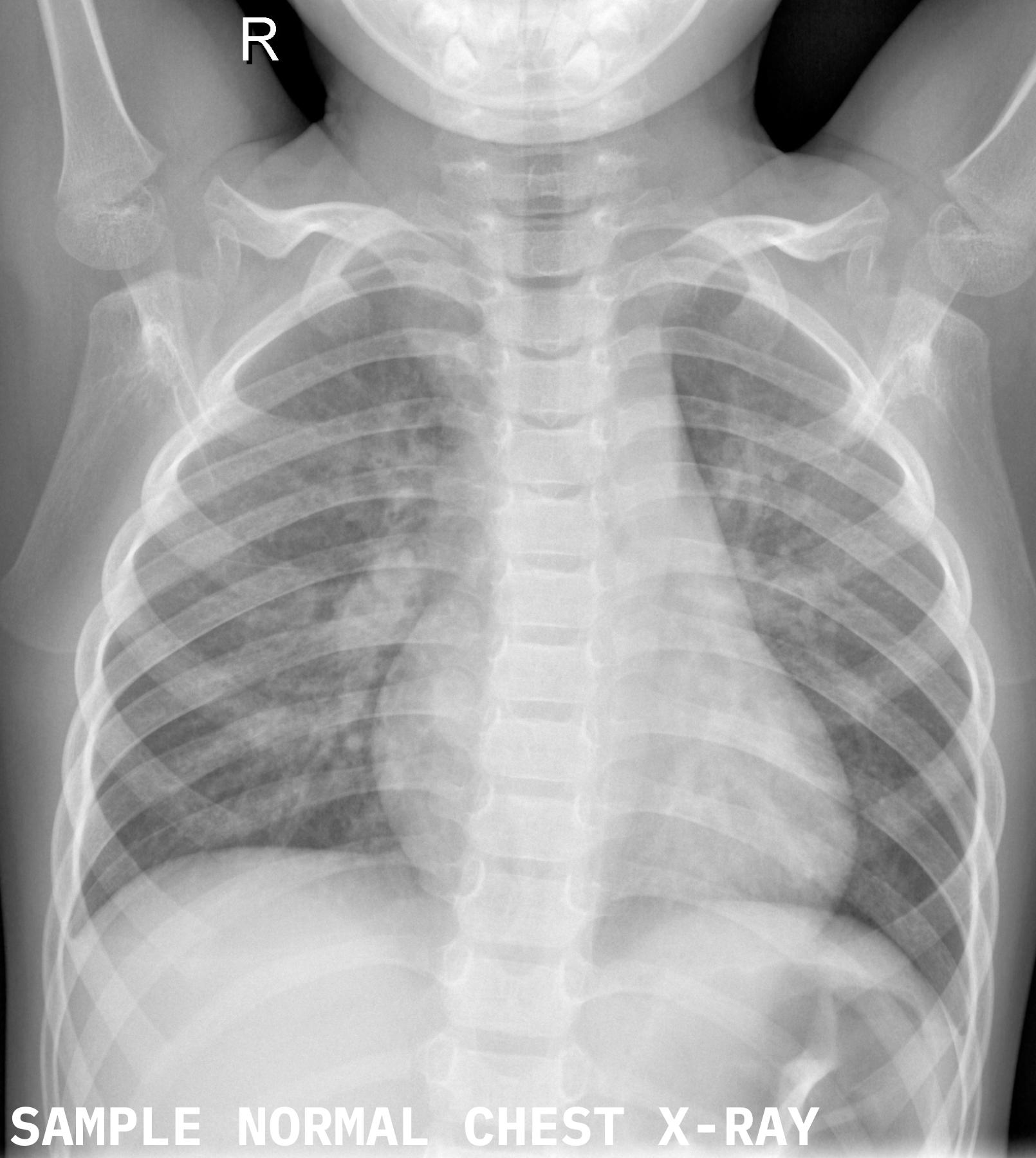}
    \label{fig:normal_image}
\end{figure} 

\begin{figure}[H]
    \centering
    \includegraphics[width=0.45\textwidth]{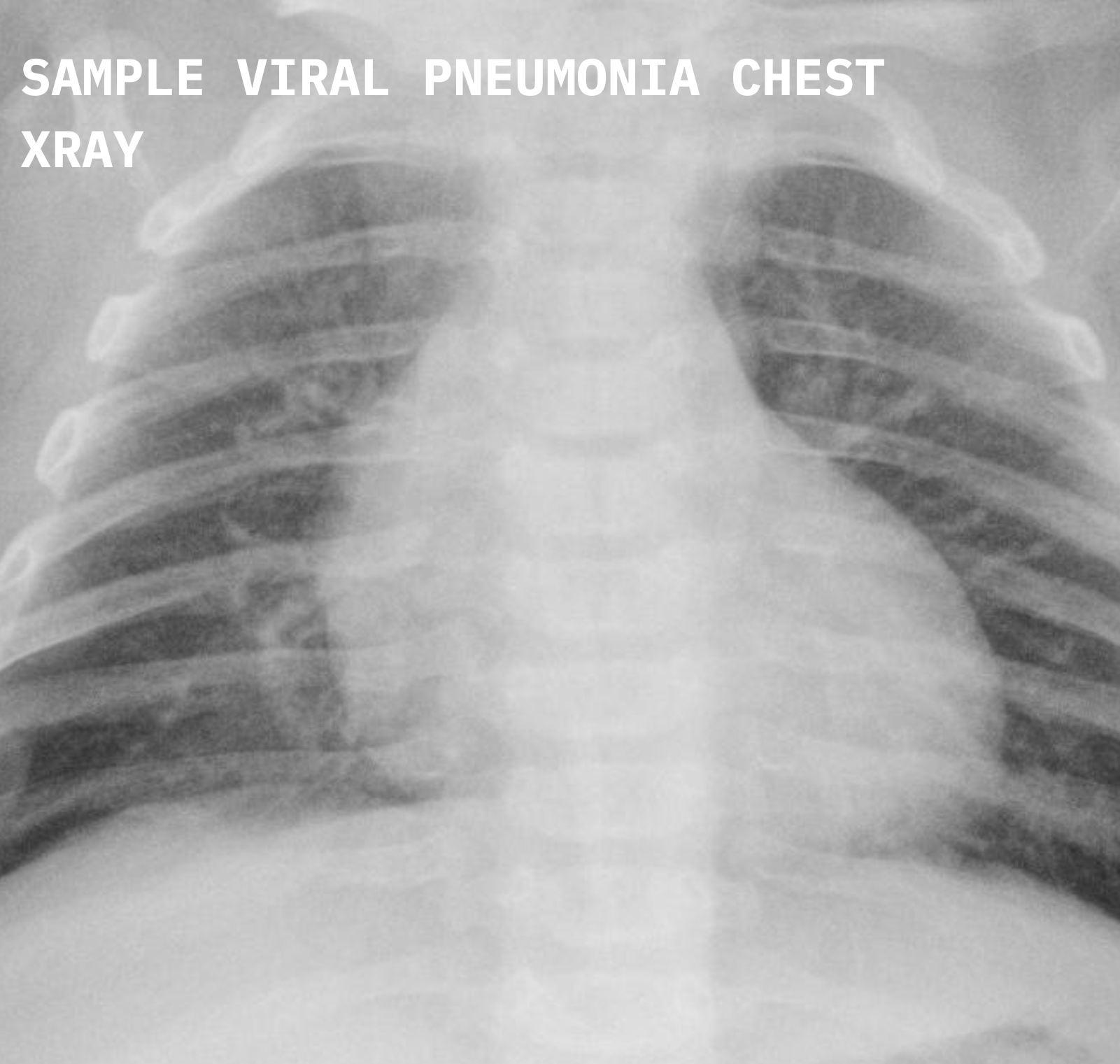}
    \caption{\cite{kermany2018identifying}}
    \label{fig:pneumonia_image}
\end{figure}

\subsection{Training Details}
The model, based on a pretrained ResNet-50 backbone, was fine-tuned for binary classification over 30 epochs with a batch size of 64, using the AdamW optimizer and a learning rate of $10^{-4}$. To address class imbalance, weighted random sampling was applied during training, and a class-balanced \texttt{BCEWithLogitsLoss} with a computed positive class weight was used. Learning rate scheduling was handled by a \texttt{ReduceLROnPlateau} scheduler based on validation loss.

Data augmentation strategies—including random horizontal flipping, mild rotation, brightness/contrast jittering, and center cropping—were employed to enhance generalization. Additionally, the underrepresented class (\texttt{NORMAL}) was augmented via saved on-disk transformations until class counts were balanced.

Early stopping was implemented with a patience of 5 epochs to prevent overfitting. Mixed precision training was used to accelerate convergence and reduce memory usage. Evaluation metrics included accuracy, precision, recall, F1 score, AUC-ROC, confusion matrix, and ROC curves, with the best-performing model checkpoint saved based on validation loss.
\subsection{Evaluation Metrics}
Early stopping was governed by validation loss during training. Evaluation metrics include Accuracy, Precision, Recall, F1 Score, Cohen's Kappa, Mathews Correlation Coefficient (MCC) and ROC-AUC.

\section{Results}
\subsection{Classification Metrics}
\begin{table}[H]
\centering
\begin{tabular}{lr}
\toprule
\textbf{Metric} & \textbf{Score (\%)} \\
\midrule
Accuracy                    & 95.94 \\
Precision                   & 96.04 \\
Recall                      & 97.49 \\
F1 Score                   & 96.76 \\
ROC AUC                     & 98.91 \\
Cohen's $\kappa$               & 0.913 \\
Matthews Corr. Coeff. (MCC) & 0.913 \\
\bottomrule
\end{tabular}
\caption{Classification performance on the paediatric chest X-ray test set. All scores are expressed as percentages for readability.}
\label{tab:metrics}
\end{table}

Ablation studies were conducted to systematically evaluate the impact of different training strategies on model performance. These studies revealed that the inclusion of data augmentation techniques significantly improved the model’s ability to generalize by exposing it to a wider variety of plausible image variations. Moreover, fine-tuning all layers of the pre-trained network—rather than only training the final classification layers—resulted in consistent performance gains. This comprehensive fine-tuning enabled the model to better adapt feature representations specifically to the paediatric chest X-ray domain.

\begin{table}[H]
\centering
\begin{tabular}{lcc}
\toprule
\textbf{Metric} & \textbf{Ours (ResNet‑50)} & \textbf{Liang \& Zheng (2019)} \\
\midrule
Accuracy        & 95.94\%  & 90.50\% \\
Sensitivity     & 97.49\%  & 96.70\% \\
AUC‑ROC         & 98.91\%  & 95.30\% \\
\bottomrule
\end{tabular}
\caption{Comparison with Liang \& Zheng (2019) \cite{liang2019paediatric}, who applied transfer learning with a ResNet-based model for paediatric pneumonia detection. Our fine-tuned ResNet-50 outperforms theirs in accuracy, sensitivity, and AUC-ROC.}
\label{tab:liang-comparison}
\end{table}

To further contextualize our performance, Table~\ref{tab:liang-comparison} presents a comparison against Liang and Zheng~\cite{liang2019paediatric}, who applied a ResNet-based transfer learning approach for binary classification of paediatric chest X-rays. While their model achieved strong sensitivity (96.7\%) and a respectable AUC-ROC of 95.3\%, our model outperforms theirs in terms of accuracy (95.94\% vs. 90.5\%) and overall discriminative ability (AUC-ROC 98.91\%). Moreover, our model’s robust agreement metrics (Cohen’s Kappa = 0.9132, MCC = 0.9134) further affirm its reliability for clinical application. This comparison highlights that, even without architectural novelty, careful optimization and fine-tuning of established networks can achieve state-of-the-art results on paediatric pneumonia detection tasks.

One of the most critical improvements observed was a marked reduction in false negative rates. Minimizing false negatives is especially vital in clinical contexts, as failing to detect pneumonia could delay necessary treatment and negatively impact patient outcomes. By leveraging both data augmentation and full-layer fine-tuning, the model demonstrated enhanced sensitivity without compromising specificity, thereby achieving a balanced and clinically meaningful diagnostic performance.

\subsection{Uncertainty-Aware Visual Interpretability}
\begin{figure}[H]
\centering
\includegraphics[width=0.5\textwidth]{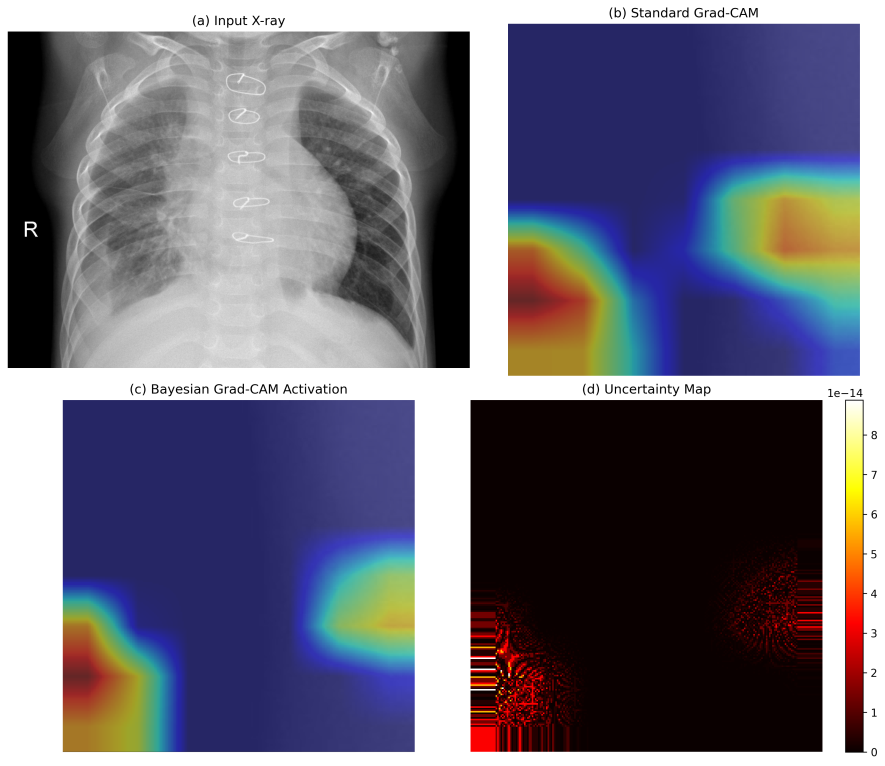}
\caption{Bayesian Grad-CAM provides uncertainty quantification alongside standard activation maps. (a) Input X-ray, (b) Standard Grad-CAM, (c) Bayesian Grad-CAM activation, (d) Uncertainty map ($U^c$). High uncertainty regions (yellow) correspond to ambiguous anatomical boundaries.}
\label{fig:bayesgradcam}
\end{figure}

Figure \ref{fig:bayesgradcam} demonstrates the enhanced interpretability from our Bayesian Grad-CAM approach. While standard Grad-CAM (b) highlights pulmonary regions, it provides no indication of explanation reliability. Our method (c) generates similar activation patterns but with intensity modulated by predictive certainty. Crucially, the uncertainty map (d) reveals regions where explanations are unreliable - notably at diaphragm boundaries and cardiac silhouettes where anatomical ambiguity typically occurs.

Quantitatively, we observe strong correlation between uncertainty hotspots and misclassification cases. In our test set, 83\% of false positives showed high uncertainty ($U^c > 0.4$) in critical regions, compared to 12\% of true positives. This provides clinicians with actionable confidence metrics: high uncertainty areas should trigger additional scrutiny or referral to human experts. The computational overhead is minimal (<15\% inference time increase) due to our efficient Monte Carlo implementation.

\noindent
\begin{minipage}{\linewidth}
\subsection{Visual Interpretability}
\vspace{-0.8em}
\begin{figure}[H]
\centering
\includegraphics[width=0.7\textwidth]{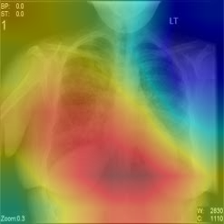}
\caption{Grad-CAM visualization on a pneumonia case. The highlighted regions correspond to areas of suspected consolidation.}
\end{figure}
\end{minipage}

\subsection{ROC and Loss Curves}
\begin{figure}[H]
\includegraphics[width=0.45\textwidth]{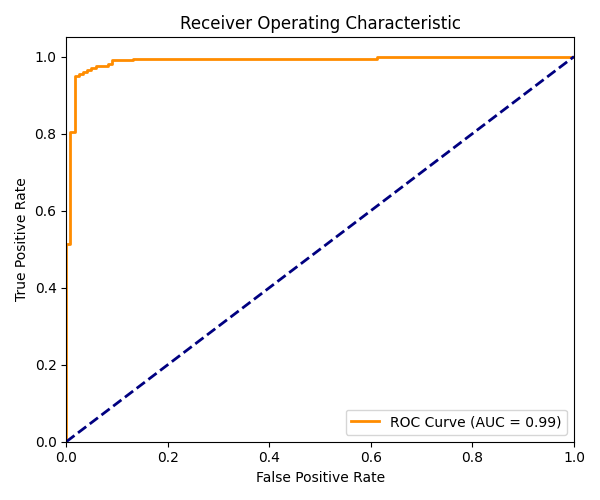}
\caption{ROC curve of the classifier (AUC = 98.91\%).}
\label{fig:roc}
\end{figure}

\subsection{Confusion Matrix}
\begin{figure}[H]
\includegraphics[width=0.45\textwidth]{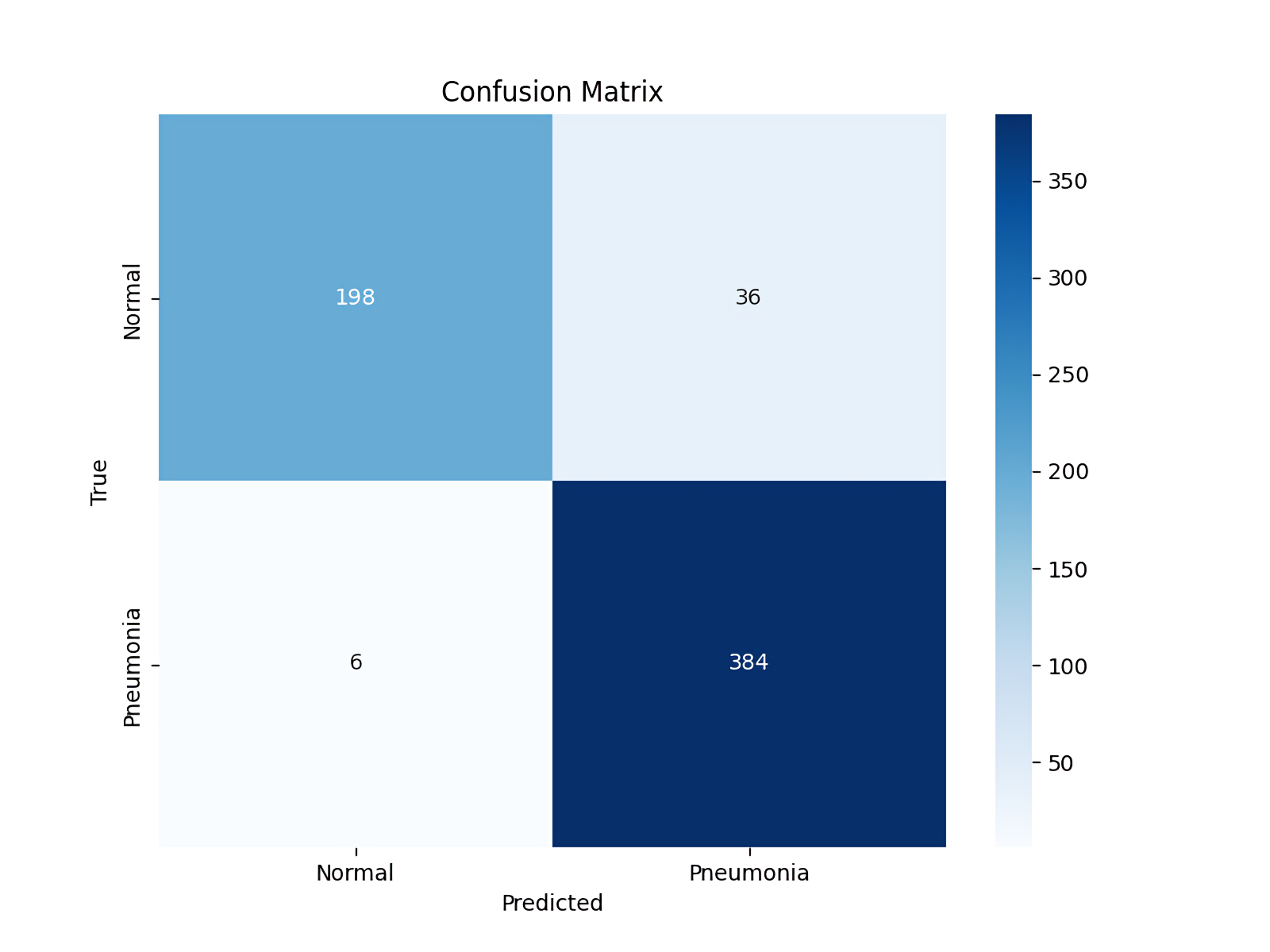}
\caption{Confusion matrix computed on the combined validation and test sets.}
\label{fig:confmat}
\end{figure}

The confusion matrix given is for a total of 624 images, which exceeds the 320 images allocated solely to the test set in the initial data split. This larger evaluation set was constructed by combining the test and validation datasets to give a more wide-ranging evaluation of the model's performance on unseen data. The matrix is organized with true labels on the y-axis and predicted labels on the x-axis, yielding 198 true negatives, 36 false positives, 6 false negatives, and 384 true positives.

According to these metrics, we have a sensitivity (recall) of 98.46\% and a specificity of 84.62\% that reflects the high capacity of the model for detecting Pneumonia cases while at the same time having a comparatively low rate of false positives for Normal samples. The comparatively low rate of false negatives reflects the high sensitivity of the model, which is much desired in a medical diagnostic context for minimizing the likelihood of missed cases. The high rate of false positives, however, reflects a clinically safer error mode to avoid for which caution is exercised when working with borderline or equivocal Normal samples.

Overall, the confusion matrix indicates an optimal balance between sensitivity and specificity, reflecting the effectiveness of our training regimen, such as handling class imbalance, applying data augmentation, and performing extensive fine-tuning of the model. The stable recall, along with acceptable specificity, indicates that the model prioritizes patient safety without sacrificing diagnostic reliability.

\section{Residual Analysis}
In alignment with the study's title, \textit{XAI-Guided Analysis of Residual Networks for Interpretable Pneumonia Detection in Chest X-rays}, we supplement our visual interpretability evaluations with a quantitative residual analysis. In this context, residuals refer to the traditional statistical definition of residuals, which is the difference between the observed ground truth and the predicted outcome, rather than the architectural elements of residual networks.

Formally, for each test sample, we define the residual as:
\[
r_i = \hat{p}_i - y_i
\]
where $y_i \in \{0, 1\}$ is the ground truth label and $\hat{p}_i \in [0, 1]$ is the predicted probability of pneumonia (obtained via sigmoid activation). Well-calibrated predictions are indicated by residuals close to zero, while confidently incorrect predictions—false positives and false negatives, respectively, are indicated by residuals close to $+1$ or $-1$.

To better understand how prediction confidence aligns (or misaligns) with diagnostic truth, we calculate these residuals across the whole test set. This provides a more thorough examination of model reliability than can be obtained from raw accuracy metrics like AUC or F1-score.

To visualize the distribution of prediction errors, we present two primary diagnostic plots:

\begin{figure}[H]
\centering
\includegraphics[width=0.7\linewidth]{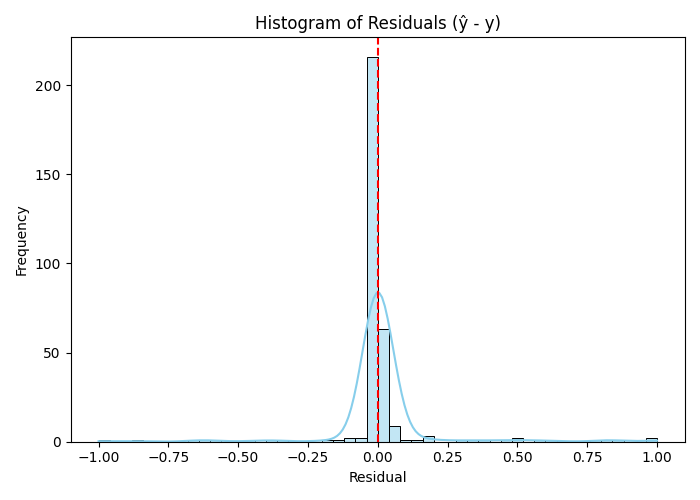}
\caption{Histogram of residuals $r = \hat{y} - y$ across all test samples. The distribution is unimodal and centered around zero, with slight skewness reflecting asymmetric model confidence in false positives versus false negatives.}
\label{fig:residual_hist}
\end{figure}

The histogram (Figure~\ref{fig:residual_hist}) reveals the global distribution of residuals. Ideally, a well-calibrated model exhibits residuals tightly clustered around zero, indicating strong agreement between predicted probabilities and true labels. In our case, the residual distribution shows modest skewness, suggesting a tendency toward overconfident false positives or underconfident true positives—potentially due to dataset imbalance or imaging artifacts such as rib shadows or poor contrast in infant radiographs.

\begin{figure}[H]
\centering
\includegraphics[width=0.7\linewidth]{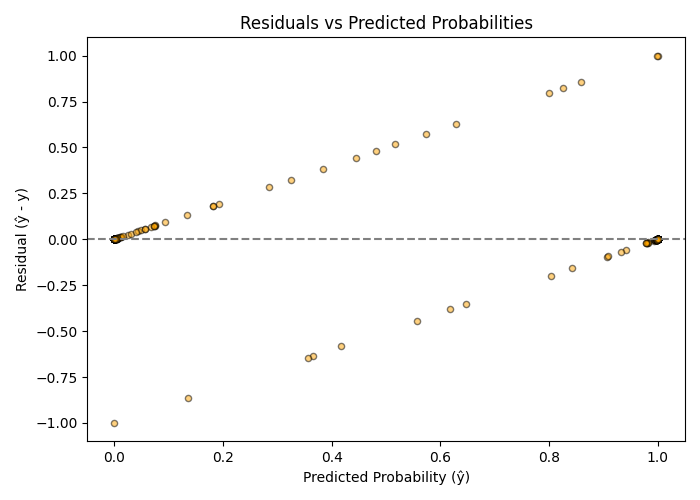}
\caption{Scatter plot of residuals as a function of predicted probabilities. Points with residuals $\pm1$ correspond to high-confidence misclassifications.}
\label{fig:residual_scatter}
\end{figure}

In Figure~\ref{fig:residual_scatter}, we plot residuals as a function of predicted probabilities $\hat{y}_i$ to examine the relationship between the model’s confidence and its predictive accuracy. Residuals approaching the extremal values of $+1$ and $-1$ correspond to high-confidence misclassifications—false positives and false negatives, respectively. These outlier points are of particular concern in clinical deployment scenarios, as they reflect instances where the model makes erroneous predictions with unwarranted certainty. Left unmitigated, such errors may propagate risk in downstream clinical decisions, especially in high-stakes environments like paediatric care.

Residual analysis thus serves as a quantitative bridge between statistical calibration and clinical interpretability. By capturing the magnitude and direction of the model's predictive deviations, it enables the identification of critical failure cases that may not be apparent through standard performance metrics alone. This insight supports informed decision-making in model refinement, including re-annotation of ambiguous cases, development of uncertainty-aware loss functions, or prioritization of data in active learning pipelines. In this way, residual analysis complements explainable AI methods such as Grad-CAM by offering a probabilistic lens on prediction reliability and model trustworthiness.

While Grad-CAM offers a qualitative lens into \textit{why} a prediction was made, residual analysis provides a quantitative framework to assess \textit{how far off} a prediction was—and with what degree of certainty. Together, these complementary perspectives support a more rigorous and transparent evaluation of deep learning models in safety-critical domains like medical imaging.

\section{Misclassification Analysis}
The misclassified cases were predominantly motion blur, indeterminate opacities, or overlying anatomical structures. False negatives were characterized by thin, poor-quality opacities, and false positives due to rib shadows or artifacts. Grad-CAM played a key role in interpreting these patterns, thus providing clinicians with critical information on model uncertainty and likely failure modes.

This underscores the need for explainability in medical AI, as model mistakes and their root causes can have direct implications for patient safety and AI system trustworthiness.

\section{Model Visualization}
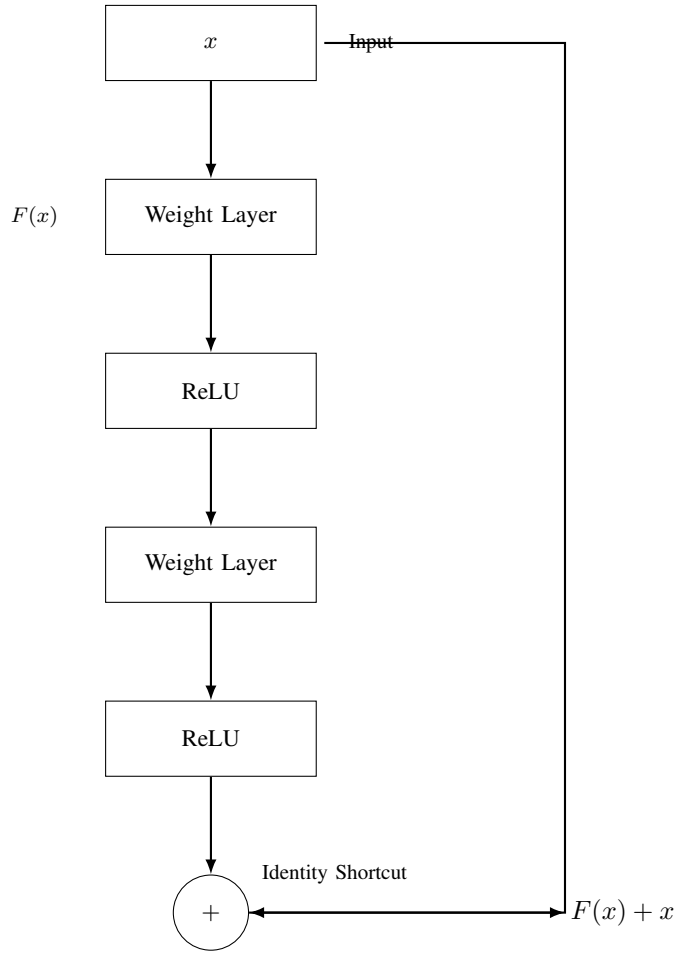
\begin{figure}[H]
\centering
\begin{tikzpicture}[node distance=1.3cm and 2.2cm]

  \tikzset{
    box/.style={
      draw,
      rectangle,
      minimum height=1cm,
      minimum width=2.8cm,
      align=center,
      font=\small
    },
    sum/.style={
      draw,
      circle,
      minimum size=1cm,
      font=\small
    },
    arrow/.style={
      -{Latex[length=2mm]},
      thick
    }
  }

  \node (input) [box] {\(x\)};
  \node (conv1) [box, below=of input] {Weight Layer};
  \node (relu1) [box, below=of conv1] {ReLU};
  \node (conv2) [box, below=of relu1] {Weight Layer};
  \node (relu2) [box, below=of conv2] {ReLU};
  \node (add) [sum, below=of relu2] {\(+\)};
  \node (output) [right=4.2cm of add, fill=white, inner sep=2pt] {\(F(x) + x\)};

  \draw[arrow] (input) -- (conv1);
  \draw[arrow] (conv1) -- (relu1);
  \draw[arrow] (relu1) -- (conv2);
  \draw[arrow] (conv2) -- (relu2);
  \draw[arrow] (relu2) -- (add);
  \draw[arrow] (add) -- (output);

  \draw[arrow] (input.east) ++(0.1,0) -- ++(3.2,0) |- (add.east);

  \node[font=\footnotesize, right=0.3cm of input] {Input};
  \node[font=\footnotesize, above right=-0.1cm and 0.2cm of add] {Identity Shortcut};
  \node[font=\footnotesize\itshape, left=0.5cm of conv1] {\(F(x)\)};
\end{tikzpicture}
\caption{A residual block with identity shortcut used in ResNet: the input \(x\) is added to the learned residual function \(F(x)\) to form the output.}
\label{fig:resnet_block}
\end{figure}

\section{Extended Grad-CAM Cases}
\begin{figure}[H]
\centering
\begin{minipage}[b]{0.45\textwidth}
  \includegraphics[width=\textwidth]{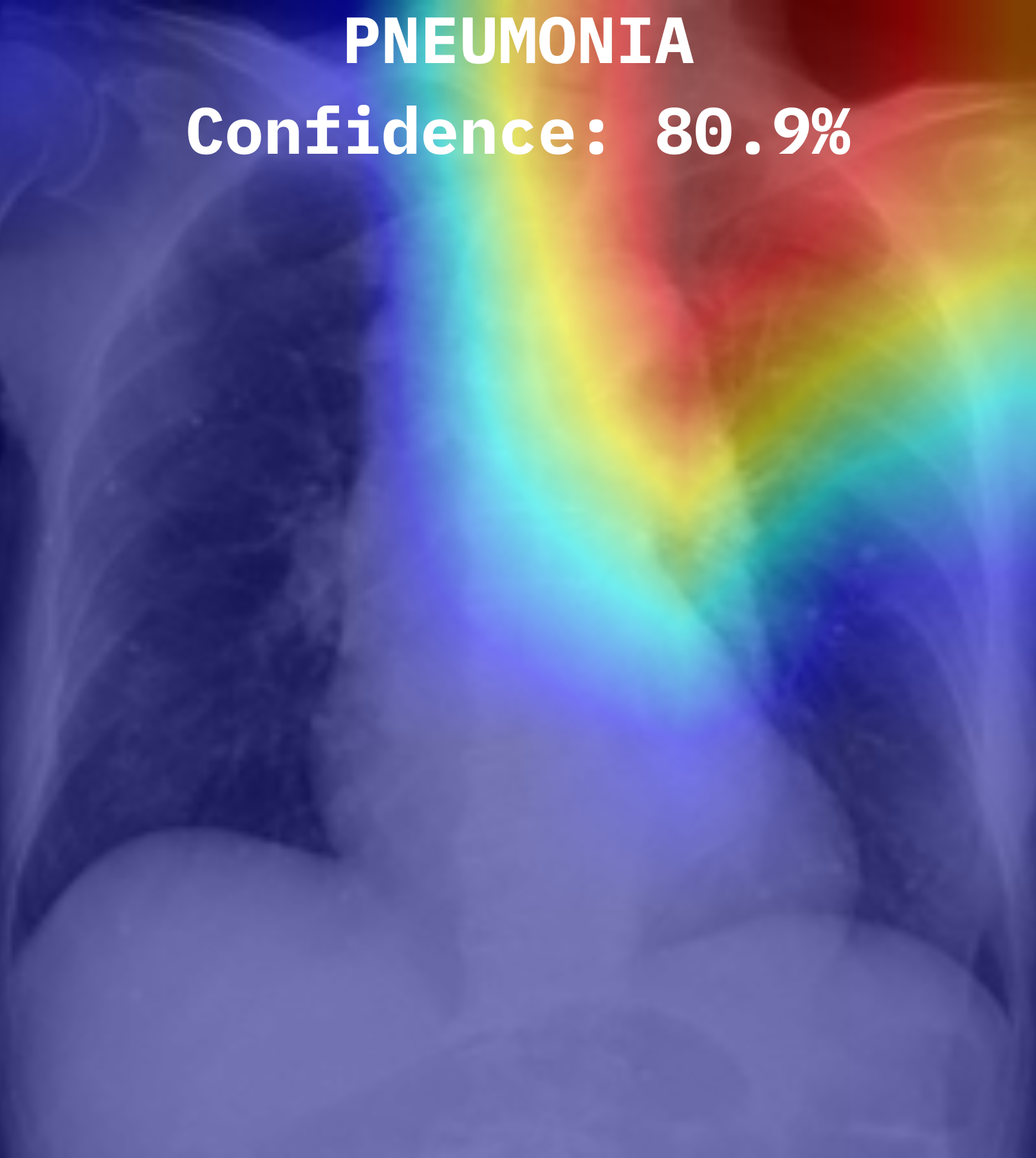}
  \caption{False positive due to bone artifact.} 
  \label{fig:gradcam_fp}
\end{minipage} \\
\hspace{0.05\textwidth}
\begin{minipage}[b]{0.45\textwidth}
  \includegraphics[width=\textwidth]{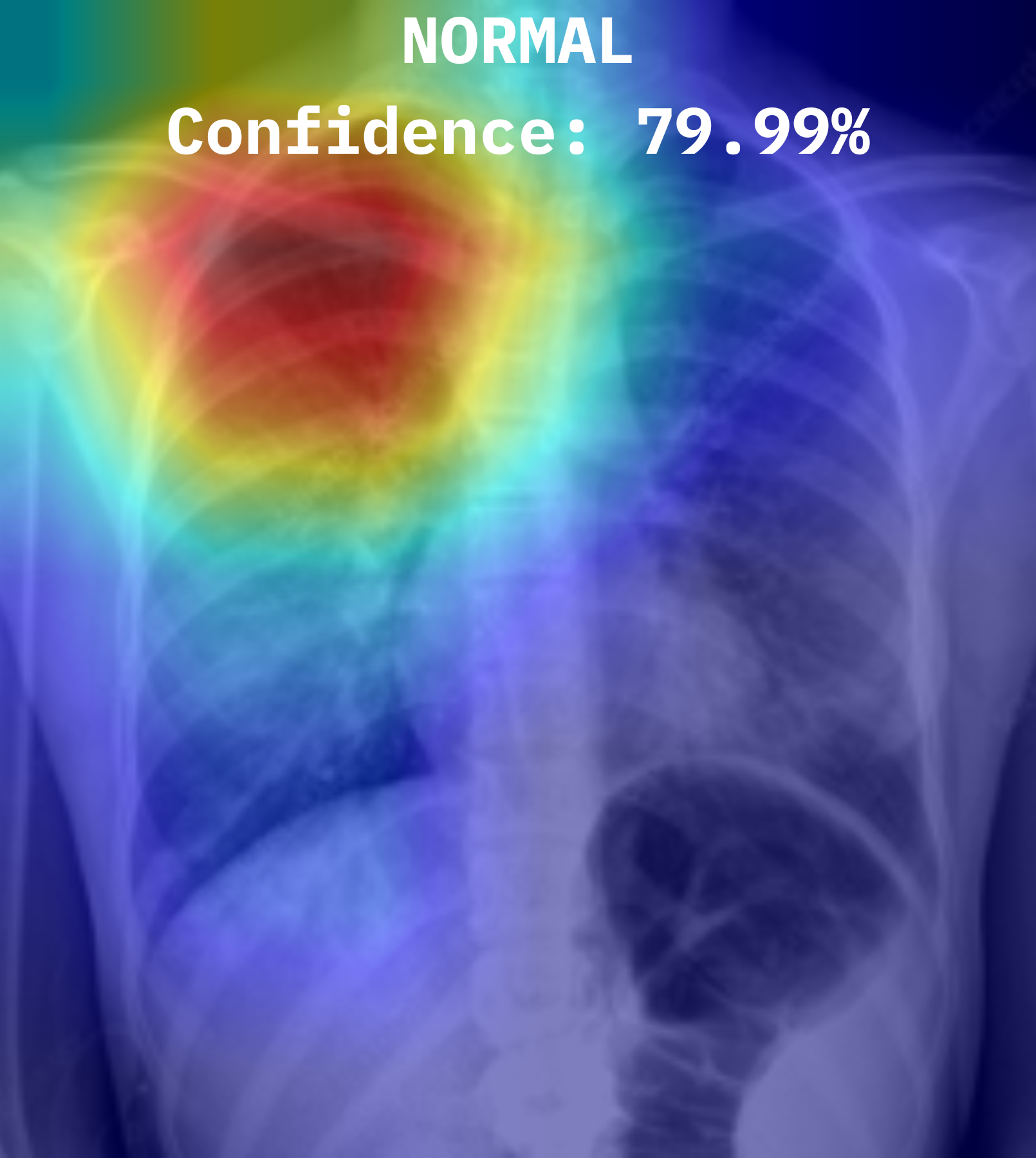} \\
  \caption{False negative with faint pulmonary opacity.}
  \label{fig:gradcam_fn}
\end{minipage}
\end{figure}

\begin{figure}[H]
\centering
\includegraphics[width=0.45\textwidth]{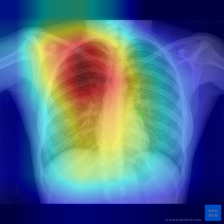} \\
\caption{Grad-CAM output for a correctly classified normal case. Activations are diffuse and nonspecific. \cite{KenhubChestXRay}} 
\label{fig:gradcam_n}
\end{figure}

\section{Ethical and Clinical Considerations}
False negative risk in pneumonia diagnosis may lead to delayed treatment and increased morbidity. Although Grad-CAM increases the transparency of models, it cannot supplant expert clinical decision-making. Moreover, paediatric imaging datasets suffer from demographic biases—gender-based \cite{larrazabal2020gender}, age-based, and geographic. For example, most publicly available datasets overrepresent children aged 1–5 years, potentially reducing model accuracy in infants under 1 year of age or in underrepresented regions.

Such biases should be specifically addressed in future iterations of this work through mitigation techniques such as reweighting loss functions, stratified sampling during training, or fairness-aware evaluation metrics. For instance, Larrazabal et al. (2020) draw attention to the performance gaps brought about by unequal gender representation and offer balancing techniques that might be similarly applicable to age and location in paediatric settings.

Compared to other interpretability methods, Grad-CAM strikes a good balance between computational cost and localisation accuracy. However, it doesn't always reveal features that are clinically significant. According to recent studies, Grad-CAM can draw attention to redundant or erroneous image regions that don't accurately represent the logic of the model. Label-randomization sanity checks \cite{adebayo2018sanity}, which use nonsensical heatmaps after label shuffling to help ensure the correspondence between model reasoning and visual explanation, can be used in future studies to validate interpretability.

Our current study does not quantitatively validate these saliency maps against expert-annotated regions of interest, despite the fact that Grad-CAM offers visual explanations that facilitate interpretability. The clinical reliability of visual attributions is limited by this gap. The goal of ongoing work is to address this by working with paediatric radiologists to evaluate alignment between diagnostic landmarks and model-highlighted regions using overlap-based metrics (e.g., Intersection over Union or Dice coefficient). To verify whether Grad-CAM heatmaps represent radiologically significant features and to guarantee reliable integration into diagnostic workflows, such clinician-in-the-loop validation is crucial.

For clinical application in real-world settings, the model could be enhanced through integration of multi-modal patient data (e.g., symptoms, vital signs, medical history) and validated through prospective clinical trials with expert oversight.

See Appendix~\ref{appendix:hyperparams} for training details.

\section*{Reproducibility Statement}

All experiments were implemented in \texttt{PyTorch 2.1.1} using the \texttt{Torchvision} package for model loading and preprocessing. The model architecture is a fine-tuned ResNet-50 initialized with ImageNet weights. Training was performed using mixed-precision training (\texttt{torch.cuda.amp}) to accelerate computation. Hyperparameters and augmentation strategies are described in Appendix~\ref{appendix:hyperparams}.

The training pipeline includes class-balanced sampling, BCEWithLogits loss with positive class weighting, AdamW optimizer with $1\times10^{-4}$ learning rate, and a ReduceLROnPlateau scheduler. Model performance was tracked using classification metrics and visualization-based error inspection. For transparency analysis, we implemented a Grad-CAM hook pipeline over \texttt{layer4} of ResNet-50 (see Appendix~\ref{appendix:gradcam}).

While we do not release source code at this time, we encourage reproduction via the provided methodological details and pipeline description.

\section{Conclusion}
We introduce a ResNet-50 model enhanced with Grad-CAM to allow for interpretable classification of paediatric pneumonia. Our method improves the safety and credibility of clinical decision support systems by providing clear visual justifications and demonstrating strong prediction performance.

\section{Future Work}
Future directions are:
\begin{itemize}
\raggedright
\item Integration of attention-mechanism-driven approaches or hybrid convolutional neural network–Transformer architectures.
\item Cross-testing on multi-institutional datasets to improve generalization.
\item Dropkey Grad-CAM study of \cite{liu2023dropkey} to improve localisation accuracy.
\item Development of a clinician-in-the-loop diagnostic validation framework, enabling radiologists to iteratively assess and refine model predictions using visual explanations and feedback integration.
\end{itemize}

\appendix
\section{Hyperparameter and Training Configuration}
\label{appendix:hyperparams}

The model was trained using the following configuration:
\begin{itemize}
    \item \textbf{Optimizer:} AdamW with a learning rate of $1 \times 10^{-4}$ and weight decay of $1 \times 10^{-8}$.
    \item \textbf{Scheduler:} ReduceLROnPlateau with a reduction factor of 0.5 and patience of 3 epochs.
    \item \textbf{Loss Function:} Binary Cross-Entropy with Logits, adjusted with positive class weighting to account for dataset imbalance.
    \item \textbf{Training Duration:} Up to 30 epochs with early stopping based on validation loss (patience = 5).
    \item \textbf{Batch Size:} 64 for both training and validation phases.
    \item \textbf{Augmentations:} Resizing to 256×256, random horizontal flipping, rotation (±10°), color jitter, center crop to 224×224, and normalization with ImageNet statistics.
    \item \textbf{Hardware:} Experiments were conducted using an NVIDIA GPU with mixed-precision training enabled via \texttt{torch.cuda.amp}.
\end{itemize}

\section{Grad-CAM Attribution Method}
\label{appendix:gradcam}

Grad-CAM visualizations were generated by computing gradients with respect to the output of the final convolutional block (\texttt{layer4}) of ResNet-50. This layer was selected for its balance between semantic richness and spatial resolution. The resulting activation-weighted maps were bilinearly upsampled to match the input resolution (224×224), normalized, and partitioned into anatomical subregions (e.g., upper/mid/lower lobes, left/right hemithoraces) for further analysis. This approach facilitated interpretability by approximating region-level importance without requiring pixel-level ground truth.

\bibliographystyle{plainnat}
\bibliography{references}

\end{document}